# An Ontological Learning Management System

Monika Rani, Kumar Vaibhav and O.P. Vyas
Indian Institute of Information Technology- Allahabad, India

ABSTRACT: The current learning systems typically lack the level of meta-cognitive awareness, self-directed learning, and time management skills. Most of the ontologically based learning management systems are in the proposed phase and those which are developed do not provide the necessary path guidance for proper learning. The systems available are not as adaptive from the viewpoint of the learner as required. Ontology engineering has become an important pillar for knowledge management and representation in recent years. The design, approach and implementation of ontology in e-learning and m-learning systems have made them more effective. In this paper, we have proposed a system for the betterment of knowledge management and representation of associated data as compared to the previously available learning management systems. Here, we have presented the application and implementation of ontological engineering methodology in the Computer Science domain. For knowledge management, we have created a domain associated ontology which represents knowledge of a single domain. Subsequently, ontology has been created to manage a learner profile so that a learner may be aligned to a proper path of learning. The learner ontology will use the VARK learning model which classifies what kind of learning does the learner requires so that necessary resources could be provided.

Keywords:Semantic Web, Ontology, e-learning, m-learning, VARK learning style, Learning management system (LMS).

## INTRODUCTION

The process of information gathering to enhance forte and the potential that can be used under various situations is known as learning. As with many evolving technology-related terms, especially "e" terms, e-Learning envelops many different things. Elliot Masie quoted: "On-line learning is not about taking a course and putting it on the desktop, it is about a new blend of resources, interactivity, performance support and structured learning activities" [1]. ELearnity has defined e-learning as the 'combination of learning services and technology to provide high value integrated learning; anytime, anyplace'[2].

The term "e-learning" was first utilized at a Computer Based Training (CBT) seminar in 1999. Other words such as "online learning" and "virtual learning" are also used as synonyms for e-learning. In the early 2000s, e-learning was presented as the next big aspect in the digital revolution. Since then, e-learning has not only removed the barrier of age, place and time but also removed the socioeconomic barriers.

For the last two decades, e-learning systems have thrived in the true sense. Even the industry has begun to use e-learning systems in order to train their employees. Employees are getting opportunities to expand their skill sets and improve upon their industry knowledge, irrespective of their experiences. The earlier e-learning systems or Learning Management Systems (LMSs) [3] were using different methods and tools to deliver various courses.

The e-learning system provides a certain amount of benefits as given below:

(1) Face to Face learning is limited to learners who have the restricted ability to participating in a particular area at some particular time. E-learning facilitates in removing such boundaries.

(2) An interactive course can be designed through the use of multimedia that enhances the engagement of the number of learners.

(3) E-learning is cost effective as it reduces the exorbitant amount of money spent to acquire updated versions of textbooks [4].

Correspondence to M. Rani (monikarani1988@gmail.com).

(4) With e-learning the professor can host a guest lecture with the help of cameras and microphones to facilitate the interaction between lecturer and learner. Learners can even replay lectures as per their convenience.

(5) The time spent in searching information is minimized by providing access to unlimited amount of resources through e-learning.

With the recent advancement in the capabilities of mobile phones along with the added advantage of inherent ubiquity, mobile devices are leveraged for learning. The improved capabilities of software and hardware, the evolving habits of mobile device users and the new advanced web browsers for smartphones have created opportunities to switch to m-learning. This method has helped in improving readiness, optimizing time management and training has also made moreaccessible for learners. As per the data provided by Ambient Insight Research, the US market has generated revenue of $958.7 million in 2010 from mobile learning products and services [5]. M-Learning can be thought of as a subset of e-Learning and the 'just enough, just in time, just for me' characteristic of m-Learning also suits the model of flexible learning as signified by Figure1.

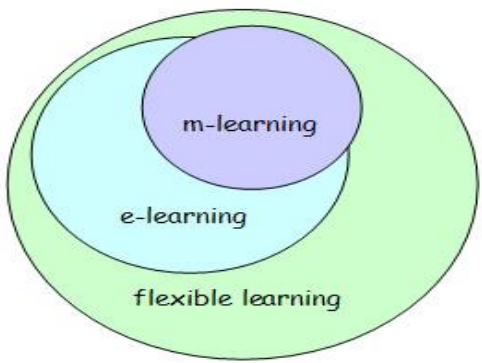

Figure 1 m-learning as a subset of e-learning

O'Malley et al. have defined mobile learning as a method where a mobile user can access the learning material wherever and whenever possible [6]. Kukulska-Hulme (2005) defined m-learning as a method where a learner can learn without being tied to any particular location. This m-learning has a feature allowing learners to be engaged in educational activities, using technology as a mediating tool, accessing data and communicating with others through wireless technology [7].

Teachers, Learners, Environment, Content, and assessment are five basic elements of m-learning. The role of teachers has slowly changed to the role of consultant in m-learning. While performing this role, teachers are required to find the learner's interest and their corresponding goals and provide them opportunities to achieve their goals. Learners are always placed at the center of each and every learning system. All other elements of the system are designed to serve learners. All parties that are part of the system must be consulted while deciding the content. Quick access to the needed information must be enabled, and therefore, it must be designed as per the pedagogical needs of learners. It must be designed in such a way so that it increases the interaction between learners as well as learner and teacher. The environment that facilitates such interactions always helps in obtaining positive learning experiences. The information needed to evaluate the learner accurately as well as formative guidance to learner is provided through attendance. An immediate motivating feedback along with assessment always motivates the learner to be an active member of the system.

The process of m-learning has some very distinguished characteristics [8]. The spontaneity in m-learning is the most defining characteristic of m-learning. The context aware m-learning has changed the e-learning into anywhere and anytime education.

(1) Privacy: A learner has access to the mobile tool and, therefore, a learner can connect and access the required content independently from other learners.

(2) Ubiquitous/Spontaneous: The spontaneity of m-learning is the most defining characteristic of m-learning. The context aware m-learning has changed the e-learning into anywhere and anytime education.

(3) Collaborative: m-learning supports interaction between learners and teachers as well as learners and learners and thus facilitates collaborative learning.

(4) Instant Information: Immediate zoning of required information is the most important characteristic of m-learning.

Learning Management System must facilitate the learner with personalization. Personalization helps the learner to identify their own learning boundaries on the basis of their preferences and personal needs. This would further help them to collaborate on the basis of information, thoughts and knowledge entities. Different Learning styles have been proposed to facilitate personalization. VARK learning style [9] is the simplest learning style as it does not involve any skill and intelligence. It is mainly focused on the method of information delivery to the learner and therefore used by most of the Learning Management Systems.

The automatic computer-based processing of abundant information on world web is done with the help of ontologydue to its machine understanding capability [10]. Information on the web needs to be supplemented with semantic knowledge with the help of ontologies. New insights can be obtained by performing inference on the knowledge represented by ontologies. Ontologies can play a major role in Learning applications if the quandary of the cognizance acquisition bottleneck is solved. A semi-automatic approach must be used to develop a domain ontology for retrieval of static and dynamic content as it reduces the cost effectiveness. Semantic knowledge is integrated to enhance the management, distribution and searching. The learning objects attested with relevant concepts forms the backbone of the ontology by creating a link between the learning material and its conceptualization. In particular, ontologies are employed in order to:

(1) Common vocabulary to share learning content
(2) Improve the reuse of learning objects
(3) The conceptualization of knowledge
(4) Personalize the learning according to his requirement
(5) Improves the efficiency of learning.

These properties of ontologies lead to vast improvements in learning management systems by defining course knowledge domain in a well defined hierarchical structure, by understanding the learning style of the learner, and by improving the assessment phase.

The rest of this paper is structured into 5 sections; initially, we focus on the introduction of the paper in section 1. In Section 2, we describe related works, where we have explained the works that have been done and proposed in the relative fields of e-learning and m-learning. Section 3 describes an adaptive learning system that makes use of the VARK learning model using an ontology to provide appropriate resources to learners on the basis of their own learning style. Section 4 depicts experimental results that prove the usefulness of the proposed an Ontology-based Learning Management System (LMS). Finally, section 5 draws the conclusion and future research opportunities in the current m-learning scenario.

RELATED WORK

Issues related to existing systems:

(1) Learning Management System: The architecture is based on the Client-Server model. The architecture consists of four main modules namely: Web Services Management, Learning Resources Management, Courseware Authoring and Graphical User Interface (GUI).
• It's hard to find ontology defined for resources other than Wikipedia.
• Stores data on the application server and hence is not scalable.
• Most of the LMS use a relational database instead of RDF/ OWL format . Thus, a database is not machine readable and we can't deploy Agent technology.
(2) We don't have standard ontology defined for online resources as well as for the users in order

to profile their data based on their learning history and interests. We build Ontology for e-learning domain due to lack of standard ontology which can fulfil the requirement of learner according to their e-learning domain (using ACM Computing Classification system (CSS), type of learner (using VARK stands for Visual, Aural, Read/write, and Kinesthetic) . As per our knowledge, there are ontology example Sweto-Dblp [23]. Sweto-Dblp ontology focused on bibliography data of publications from DBLP with additions that include affiliations, universities, and publishers. An RDF schema based on Sweto-Dblp is merely used to store bibliographic data connected with learning objects but not the resource like e-book & video etc.

(3) The VARK model determines the learner's learning type. Learning management system facilitating the learner with personalization due to VARK model which determine the learner's learning type and preference.

(4) Ontology defined is not scalable and flexible as they are deployed on the application server rather than on the cloud. We develop a scalable and flexible m-Learning System with a well-defined ontology to associate various resources stored on the cloud with each other as well as the end users. For the same we proposed Architecture for Ontology-based Learning Management System (LMS) (Figure 10). An Ontological Learning Management System is a motivational step in m-learning and Semantic web education learning (SWEL).

Basic Elements of m-Learning:

Teachers, Learners, Environment, Content and Assessment are five basic elements of m-learning. All these elements have been described below:

Teachers: The role of teachers has slowly changed to the role of consultant in m-learning. While performing this role, teachers are required to find the student's interest and their corresponding goals and provide them opportunities to achieve their goals. Following are some basic responsibilities that must be performed by teachers in m-learning system:

(1) Strengths of current methods used must be identified by teachers.
(2) Weaknesses of current methods used must be identified as well as provide alternative methods
(3) to eradicate the identified weaknesses.
(4) Must arrange interactive sessions amongst collaborative groups and thereby motivate
(5) learners.

Learners: Learners are always placed at the center of each and every learning system. All other elements of a system are designed to serve learners. Following are some basic responsibilities that must be performed by teachers in m-learning system:

(1) Accessing information as per their convenience.
(2) New information or product that must improve the system should be suggested.
(3) Should identify and use their learning styles.

Content: All parties that are part of the system must be consulted while deciding the content. Quick access the needed informationmust be enabled and, therefore, it must be designed as per the pedagogical needs of learners.

Environment: It must be designed in such a way so that it increases the interaction between learner and learner as well as learner and teacher. The environment that facilitates such interactions always helps in obtaining positive learning experiences.

Assessment: The information needed to evaluate the learner accurately as well as formative guidance to learner is provided through attendance. An immediate motivating feedback along with assessment always motivates the learner to be an active member of the system.

The past and the present of m-learning:

The e-learning timeline is shown in Table 1. The history of m-learning has been characterized by three phases as shown in Table 2. The first, second and third phases representedfrom a perspective ofdevices' focus, focus on learning outside the classroom and learner's mobility's focus respectively.

Role of ontology in m-learning:

Ontologies can play a major role in learning applications if the quandary of the cognizance acquisition bottleneck is solved. A semi-automatic approach must be used to develop a domain ontology for retrieval of static and dynamic content as it reduces the cost-effectively. Semantic knowledge is integrated to enhance the management, distribution, and searching. The learning objects attested with relevant concepts forms the backbone of the ontology by creating a link between the learning material and its conceptualization.In particular, ontologies are employed in order to:

(1) Improve the reuse of learning objects;
(2) The conceptualization of knowledge.

Ontology, the backbone of Web 3.0 [24], is used in as semantic web language processing. It lays the foundation of describing a domain of interest by describing the terms organized in a hierarchical structure that shape the reality. The concepts and the properties defining these concepts, as well as the relationship between different concepts, constitute ontology. All users of the digital space share the same vocabulary in order to facilitate intelligent processing of information as well as communication and thus laying the foundation of liaison between computers. The paper [25] describes the methodology of creating ontologies for a particular domain, such as Computer Science domain and also explains the utilization of such ontology in e-learning application. The different levels of ontology such as low-level ontology and high-level ontology have been discussed in details. The paper does not mention any ontology related to the user as well as the resources such as lecture notes, videos. The proposed system makes use of a database to handle the same details and thus fails to make use of semantic data associated with users and resources. The paper [26, 27] describes various methodologies of creating ontologies to maintain user profiles. The basic characteristics of users have been determined and are defined as concepts and the various attributes that define these characteristics are defined by various properties. In the paper [28], an ontological e-learning system has been proposed, that provides the resources to the students as per their preferences. The proposed system does not provide any specific domain ontology and an architecture that would run efficiently on mobiles. The comparison between Traditional Learning Management systems and Learning Content Management Systems (uses ontology) is given in Table 3. The comparison Table 3 clearly illustrates the added advantages that can be achieved by making use of ontologies in the learning management system for developing e-learning system.

Table 1  e-learning Timeline

| Year | Achievements |
| --- | --- |
| 1840 | Issac Pitman taught first correspondence course through the mail system. Students sent their assignments using the mail system [11]. |
| 1924 | First the machine to conduct test was invented through which students can test themselves [12]. |
| 1954 | A teaching the machine that was used to administer program instructions was invented by BF Skinner [13]. |
| 1960 | PLATO (Programmed Logic for Automated Teaching Operations) was introduced as the first automated training program [14]. |
| 1996 | Jones International University was put up as the first credited web based university [15]. |

Table 2  m-learning Timeline

| Phase | Year | Project/Scientist | Milestone |
|---|---|---|---|
| | 2003 | PERRY [16] | PDAs were used by teachers in school to evaluate |
| The First Phase: Devices' Focus | 2004 | McFARLANE [17] | Hand e-learning was introduced in Bristol University where each student and teachers were provided mobile phones. |
| | 2004 | Sharples et al. 2002 [18] [19] | HandLeR (Handheld Learning Resource) was the first web-based application used for m-learning. ADatabase manager, communication manager, tools for capturing events and a web browser were four important modules of this application. |
| The Second Phase: Focus on learning outside the classroom | 2005 | MOBIlearn [19] | First m-learning system to make use of position aware-systems |
| | 2006 | Learn eXact [20] | First m-learning system to use context-aware learning. |
| | 2007 | MyArtSpace[21] | Learners used to take notes while visiting some important location that was shared to a common web space. It was the first application that combines a learner's personal, physical and virtual space altogether. |
| The Third Phase: Learner's mobility's focus | 2007 | CAGE [22] | It uses the context's interaction model. |
| | 2008 | MARA (Nokia) [22] | Camera-equipped mobile devices were used for online video training. |

Table 3  Comparison of Various Learning Management System(LMS)

| | Traditional LMS | Traditional LCMS | Modern Learning System |
|---|---|---|---|
| **LEARNING EXPERIENCE** | | | |
| Learning Interventions | Formal | Formal, Informal | Formal, Informal, and Social |
| Learner Engagement | Classroom, Desktop | | Classroom plus on-demand learning via web or mobile: anytime, anywhere and on any device. |
| End User Tools | Browse Catalog, Simple Search, Email notification | | Browse catalog, Faceted search, Individual Development Plan, Dynamic Recommendations, Learning Paths, Learner/Manager Dashboards, Email/Text Notifications, Ratings, and Reviews, Badges/Leaderboards |
| **LEARNING CONTENT** | | | |
| Content Authoring | Simple Course | Sophisticated and | Granular learning content |
| | Builder, Rapid Authoring Tools | structured authoring for courses, presentations, printed guides, job aids, web pages, and Flash | separated from a presentation for rapid assembly and reuse across any output format or audience |
| Content Management | | Versioning, workflow, and review tools | Online collaboration, versioning, workflow, and review tools |
| Content Publishing | | Print and web output templates/formats | Print, responsive web (HTML5/CSS3),and mobile output templates/formats |
| Digital Content Delivery | Courses<br>• SCORM/AICC<br>• Instructor-led | Packaged SCORM/AICC or .PDF | Learning Object Repository (LOR) serving multiple formats for many systems and devices |
| LEARNING ADMINISTRATION | | | |
| User and Group Administration | Manage Users and Groups | | Single Sign-On (SSO),<br>Integrated with HR system |
| Course Administration | Course Enrolment, Completion Rules, Classroom Management | | Course Enrolment, Learning Paths, Classroom Management, Competency Management, Certification Management |
| Reporting and Analytics | Completion Tracking, Test Scores | | Centralized Learning Record Store (LRS) for reporting and analytics, Completion Tracking, Test Score, Question Analytics, Informal Learning Activities, Social Learning Activities, Content Effectiveness |

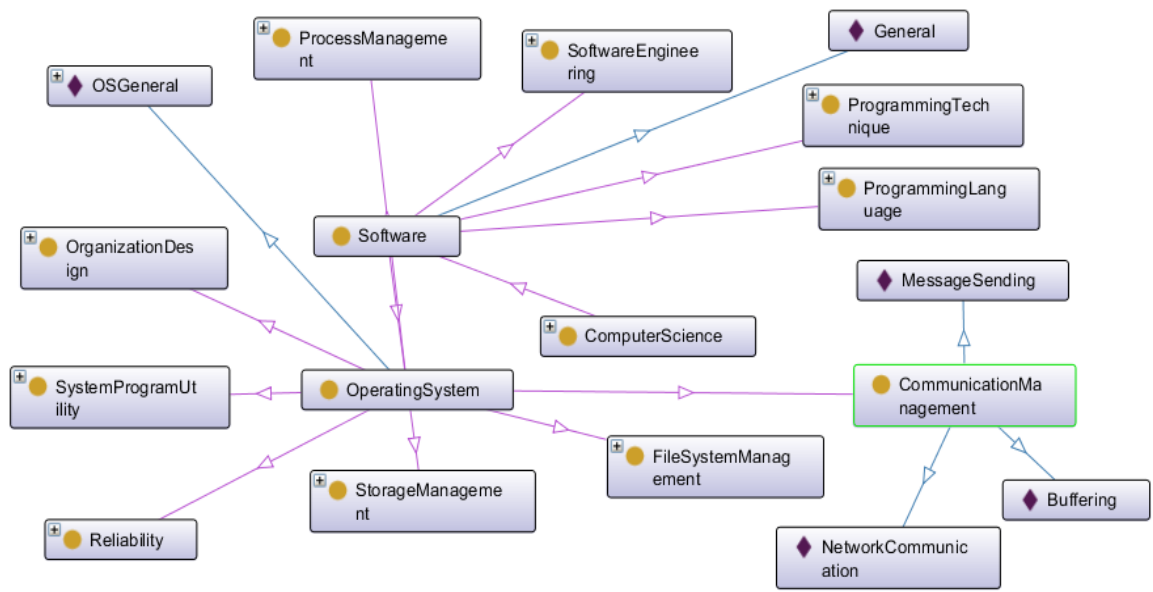

Figure 2 Domain ontology using ACM Computing Classification System (CSS)

The author R. Robson provided an exhaustive work on e-learning by describing the history of e-learning and its evolution [29]. The elementary unit of learning called the learning objects (LOs) represents the content in the e-learning system. The three layers of modeling the e-learning system were studied by Siqueira et al. [30]. The three layers, thus described are semantics embedded in LOs, representation of LOs and learning activities and the third layer for achieving the learning objectives due to coordination of the content.

An example of ontology-based learning is CRUONTO Ontology [31, 32]. It is designed for curriculum management along with program review and assessment. Using ontology based learning has various advantages that include management of resources, sharing of resources; a learner can personalize the learning according to his requirement, and improves the efficiency of learning. The CURONTO ontology reviews the curriculum, decision making, finds the discontinuity in learning and removes the redundancy in the curriculum.

A knowledge-based application that maintains the complex relationship between the different content of a single curriculum is provided under the Crampon project [33]. It involves inquiry-based learning where the elements of data and their entities are identified. K. Litherland and D. Castellanos-Nieves describe ontology based learning method, ontology E-learning (OeLe) [34, 35]. It marks the learner answers to the questions having conceptual nature. Apart from this it also provides a platform for feedback for learners and lecturers. An AI system is used to mark the learner's answers. It includes both open and closed answers. A. Sameh emphasizes on ontology-based feedback e-learning [36]. A medically based learning ontology in [37] is a domain-based learning process and is known as Unified Medical Language System (UMLS). It generates the hints to assess the learner's answer to any question. It helps in guiding a learner towards proper learning.

Ontology for mathematical learning is presented in RAMSys [38]. It corrects the mathematical formulae semantically. It also uses the taxonomy of the mathematical symbols to provide feedback. To search for higher education courses in [39], CUBER project (Curriculum Builder in the Federated Virtual University of the Europe of Regions) was produced. Under this project, the learners provide his educational information and using it the CUBER project identifies the best course. It incorporates the LOM standard of E-Learning.

To represent the e-learning framework in [40], a model is used which represents stores and retrieves the domain knowledge. It makes use of manufacturing concepts, learning concepts and also uses context. For formation of any new product, it uses a standardization model. To search and to retrieve any information, it uses the existing rules and structures. The Information Workbench presented in [41] provides a platform to collect, store, edit, report and analyze any of the available learning resources. The ontology presents and structures the available course contents for a better learning experience.

From the viewpoint of student's/learner's academic progress, a model is presented in [42]. It is based on student's interest and personal characteristics possessed by them. The characteristic of a student includes a level of understanding towards a particular subject, interest in the field of study, preference towards the interaction process and learning goals. Also, author Paneva-Marinova [43], presents a student's model of e-learning. Under this model, the ontology is comprised of two major parts. These are the student's basic information and inclination towards learning. For disability aware learning, ONTODAPS system was introduced in [44]. It is for both the students; with or without the disability. An appropriate learner level is allowed which personalizes the resources for learning. It is an e-learning which relates the learning procedure to personalize the educational process. Its framework is web service based; learning units have semantic information and the relationship between the learning the resources. A curriculum construction which can be personalized with a student has been based on an appropriate learning object combination which was introduced as PASER (Planner for the Automatic Synthesis of Educational Resources) [45]. Every learner has a profile and own preferences, which are used to design a course for personalized study. This system consists of three parts which are: a metadata repository, a system based on deductive object orientation and HAPEDU which is a planning system which constructs the curriculum.

In 2013, Cisco has expanded the definition of IoT to the Internet of Everything (IoE), which includes people, places, objects and things. Basically, anything you can attach a sensor and connectivity can participate in the new connected ecosystems. The Smartphone can be connected to college Sensor data and then the analysis can be done on that data [46].

The GroupeSpeciale Mobile Association (GSMA) report in "The Mobile Economy 2014" That the Total connected Device: 11.3 Bn in 2013 and expected to connect in 2020 is 25.7 Bn in 2020. Also Mobile connected Devices: 6.9 Bn in 2013 and expected to be 10.8 Bn in 2020 [47].

Ontologies and Linked Open Data integrate heterogeneous sources and allow federated query using mobile application for various purposes like m-learning, information retrieval (Ontology-based semantic question answering system) [48], health domain, etc. [49].

## PROPOSED ONTOLOGY- BASED LEARNING MANAGEMENT SYSTEM (LMS)

The ontology has been classified in two categories: domain ontology [50] and task ontology [51]. The domain based ontology has been developed by making use of ACM Computing Classification system (CSS) [52]. This ontology classifies the different technical terms used in different subjects by an organized hierarchical structure. In the proposed an Ontological Learning Management System (LMS), ontologies are defined, varying with their objectives. The upper-level ontology is defined in computerScience.owl file that defines the basic hierarchy as described in the ACM Classification of Computer Science field. This ontology classifies the different technical terms used in different subjects by an organized hierarchical structure. Figure 2 show the classification of the Computer Science field. The Software is a direct subclass of Computer Science and consists of Software Engineering, Operating system, Programming Language, and Programming Technique as subjects. These subjects such as Operating systems are again classified into various sub-topics such as Process Management, Reliability, and CommunicationManagement, etc. The final topics such as Message Sent, NetworkCommunication, and Buffering are described as individualsor instances of CommunicationManagement. Figure 3 show snippet code use for the declaration of Buffering as an individual of CommunicationManagement which in turn is declared as a subclass of the OperatingSystem.The other subjects and the topics associated with them are added to the domain ontology in a similar way. The task-based ontology deals with users and resources.

```
<SubClassOf>
<ClassIRI="http://www.semanticweb.org/vaibhav/ontologies/computerScience#CommunicationMgt/>
<ClassIRI="http://www.semanticweb.org/vaibhav/ontologies/computerScience#OperatingSystem/>
</SubClassOf>

<ClassAssertion>
<NamedIndividual IRI="#Buffering"/>
</ClassAssertion>
```

Figure 3 Snippet shows declaration of Buffering as an individual of CommunicationManagement

The other ontology is defined in userResource.owl file which contains the details of resources such as videos, lecture-notes, books, etc. These resources consist of data properties [53] that link files placed on the application server to the individuals defined under a particular resource in computerScience.owl ontology file. For example, the CommunicationManagement Resource is defined as an individual under the Lecture Notes which in turn is described as a subclass of resouce field as shown in Figure 4.

The other important class of this ontology is the class that represents the various end users that are using this system. These users include student, teachers and admin and these are defined as subclasses of the user class. These users have specific data property such as a user id that is used to identify them. Figure 5 representing an individual student and an individual teacher snippet.

```
<SubClassOf>
<ClassIRI="http://www.semanticweb.org/vaibhav/ontologies/userResource#LectureNotes/>
<ClassIRI="http://www.semanticweb.org/vaibhav/ontologies/computerScience#Resource/>
</SubClassOf>

<ClassAssertion>
<NamedIndividual IRI="#CommunicationManagementResource"/>
</ClassAssertion>

<DataPropertyRange>
<DataProperty IRI="#path"/>
<Datatype abbreviatedIRI="xsd:string">
</DataPropertyRange>

<DataPropertyDomain>
< DataProperty IRI="#path"/>
<Class IRI="#Resource"/>
</DataPropertDomain>

<DataPropertyAssertion>
<DataProperty IRI="#path"/>
<NamedIndividual IRI="#CommunicationManagementResource"/>
<Literal datatypeIRI="&xsd;string">localhost:8080/thesisMLearning/CMResource.pdf</Literal>
</DataPropertyAssertion>
```

Figure 4 Description about userResource.owl and computerScience.owl

```
<DataPropertyAssertion>
<DataProperty IRI="#userid"/>
<NamedIndividual IRI="#abcStudent"/>
<Literal datatypeIRI="&rdf;PlainLiteral">abc05@gmail.com</Literal>
</DataPropertyAssertion>

<DataPropertyAssertion>
<DataProperty IRI="#name"/>
<NamedIndividual IRI="#xyzTeacher"/>
<Literal datatypeIRI="&rdf;PlainLiteral">XYZ Teacher</Literal>
</DataPropertyAssertion>
```

Figure 5 Snippet for representing individual student and teacher

After identification of individuals and their respective classes, various object properties are defined to interconnect individuals with each other. ABC Student is a student of XYZ Teacher or we can say that XYZ Teacher is teacher of ABC Student and the same relationship holds for each and every student and teacher. Such object properties are also known as Inverse Object Properties. Therefore, these properties are defined as object properties in our ontology as describe in Figure 6.

```
<ObjectPropertyDomain>
<ObjectProperty IRI="#studentOf"/>
<Class IRI="#Student"/>
</ObjectPropertyDomain>

<ObjectPropertyRange>
<ObjectProperty IRI="#studentOf"/>
<Class IRI="http://www.semanticweb.org/abhishek/ontologies/userResource#Computer!
</ObjectPropertyRange>

<InverseObjectProperties>
<ObjectProperty IRI="#teacherOf"/>
<ObjectProperty IRI="#studentOf"/>
</InverseObjectProperties>

<ObjectPropertyAssertion>
<ObjectProperty IRI="#studentOf"/>
<NamedIndividual IRI="#abcStudent"/>
<NamedIndividual IRI="#xyzStudent"/>
</ObjectPropertyAssertion>
```

Figure 6 Snippet for defining object properties

The various individuals identified in computerScienceOntology are linked to the individuals defined in the resource class as well as the user class. The resource such as CMResource.pdf is associated with the individuals MessageSending, NetworkComunication and Buffering of Communication management class which is a subclass of Operating System class. The object property "contains" links all the individuals of Communication Management with the CMResource.pdf. The Figure 7 snippet represents the same association.

```
<ObjectPropertyDomain>
<ObjectProperty IRI="#contains"/>
<Class IRI="#Resource"/>
</ObjectPropertyDomain>

<ObjectPropertyRange>
<ObjectProperty IRI="#contains"/>
<Class IRI="#CommunicationManagement"/>
</ObjectPropertyRange>

<ObjectPropertyAssertion>
<ObjectProperty IRI="#contains"/>
<NamedIndividual IRI="#CMResource"/>
<NamedIndividual IRI="#MessageSending"/>
</ObjectPropertyAssertion>

<ObjectPropertyAssertion>
<ObjectProperty IRI="#contains"/>
<NamedIndividual IRI="#CMResource"/>
<NamedIndividual IRI="#Buffering"/>
</ObjectPropertyAssertion>
```

Figure 7 Snippet for representing association between resource and individuals

This ontology contains each and every detail associated with the learners, including the learning style that would suit the most. VARK learning model has been used in the proposed system. To implement the VARK Learning model, a data property known as VARK is created and the various types of learners, such as visual, aural, read/write and kinesthetic learners are assigned as values to VARK data properties. The learner is asked a set of questions while registering with the system. Based on the answers given by the learner, the system will be able to determine the type of learner as visual, aural or others. The other important attributes such as date of birth, contact number, etc. are added for each type of learners as data properties. The other class contains the details of resources such as videos, lecture-notes, books, etc. These resources consist of object properties that link the files placed on the server to the individuals defined under a particular resource in domain related ontology file. For example, the ProcessManagement Resource is defined as an individual under the Lecture Notes which in turn described as a subclass of resource field. These resources and the learners/users/students are associated with the individuals of classes defined in the domain ontology in order to provide or recommend the correct resources to the learners as shown in Figure 8. Similarly, a student is associated with the teacher using the 'isStudentOf' object property. The various important object properties associating different individuals are shown in Table 4.

Table 4 Object Properties

| Name | Domain | Range | Inverse |
|------|--------|-------|---------|
| IsStudentOf | Student | Teacher | isTeacherOf |
| IsTeacherOf | Teacher | Student | isStudentOf |
| Teaches | Teacher | Computer Science | taughtBy |
| taughtBy | Computer Science | Teacher | Teaches |
| UploadedBy | Resource | User | Not Applicable |
| IsPursuing | Student | Computer Science | enrolledAt |
| enrolledAt | Computer Science | Student | isPursuing |
| addedBy | User | User | Not Applicable |
| Contains | Resource | Computer Science | Not Applicable |

Another class consisting of questions and hints has been created to make a learner learn in an efficient manner. For multiple choice questions, a learner would be given the first chance to answer a question. In acase of thewrong answer, the learner would be given a hint. The wrong answer in the second attempt would recommend the learner an appropriate resource to go through.The process has been shown as a flow chart in Figure 9.

The learner goes through the following phases while using the system:

(1) Registration: This will facilitate the system to gather the general information of the learner.
(2) Survey: A questionnaire based on VARK model to identify the learning style.
(3) Accessing Resources: Based on the learning style and the courses enrolled, resources would be presented to the learner.
(4) Evaluation/Test and Feedback: A learner would be evaluated through quizzes, tests, etc. and based on performance, relevant feedbacks are provided.

The session, adaption and learner module work together to facilitate the learner.

The proposed architecture for an ontological learning management system as shown in Figure 10 consists of an application layer and an infrastructure layer. The application layer provides management features, i.e. course management, user management and site management, whereas infrastructure layer is used as a dynamic and scalable host pool. The master server distributes the requests to other slave servers based on their availability. Some of the slave servers are used as storage servers and the various ontologies such as ComputerScience.owl, user.owl and database are deployed on these servers, whereas the other slave servers are used to handle the client requests by interacting with the resources stored on the storage slave servers. We use database (Oracle 10g) to store Login information (user-id, password & type of user/learner). And OWL API is used to perform an operation such as insert, update, create and delete in an ontology. This API is stored as a resource in an application server. The external websites can make use of the ontology to extract any required information as per their requirements. The advantages of the proposed architecture are as follows:

(1) Powerful computing and storage capacity
(2) High Availability
(3) High Security
(4) Virtualization
(5) Provides easy access

Thus, the proposed system provides an adaptive learning system that makes use of the ontology to provide appropriate resources to learners on the basis of their learning style as well as recommend them proper resources to study on the basis of assessment of their quiz by VARK learning model.

Figure 11 and Figure 12 show use cases and components for our an Ontological Learning Management System system. The various end users or actors that will use in our system are:

(1) Admin
(2) Teacher
(3) Student

Admin: Admin is the super user or the power user of the system. Admin can play the role of all users and, therefore, can perform all the functionalities in the system. The important use cases or scenarios associated with admin are: Adding Student, Adding Course, Adding Teacher, Adding Manager, Deleting Teacher, Deleting Student, and Deleting Course.

Teacher: Teacher can add courses as well as update courses. A Teacher can add various resources related to different courses as well as students.

Student: Student can join any course as well as view and update his profiles.

Manager: Manager can update courses as well as upload various resources related to different topics that are associated with the course

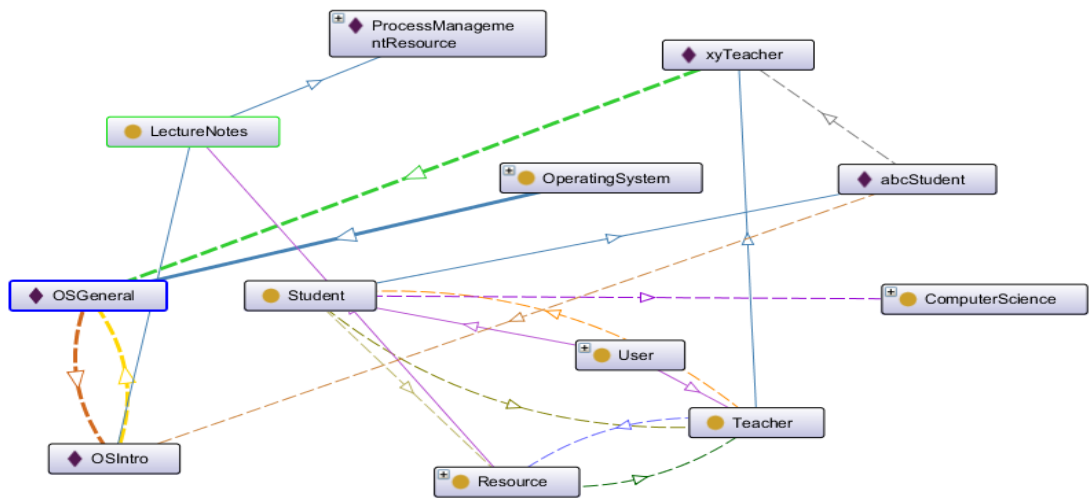

Figure 8 Association of Resources with learners/users

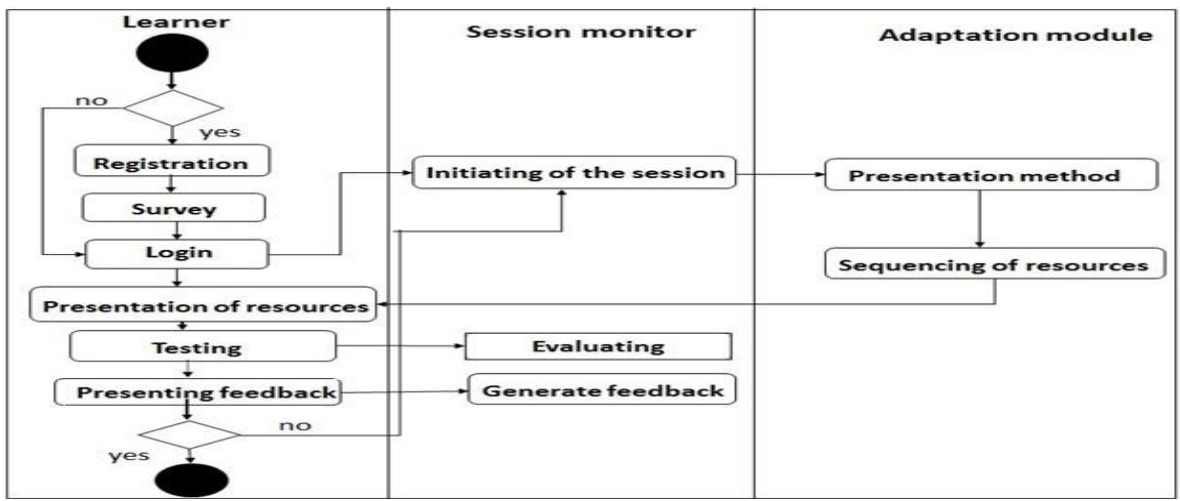

Figure 9 Workflow from learner's perspective

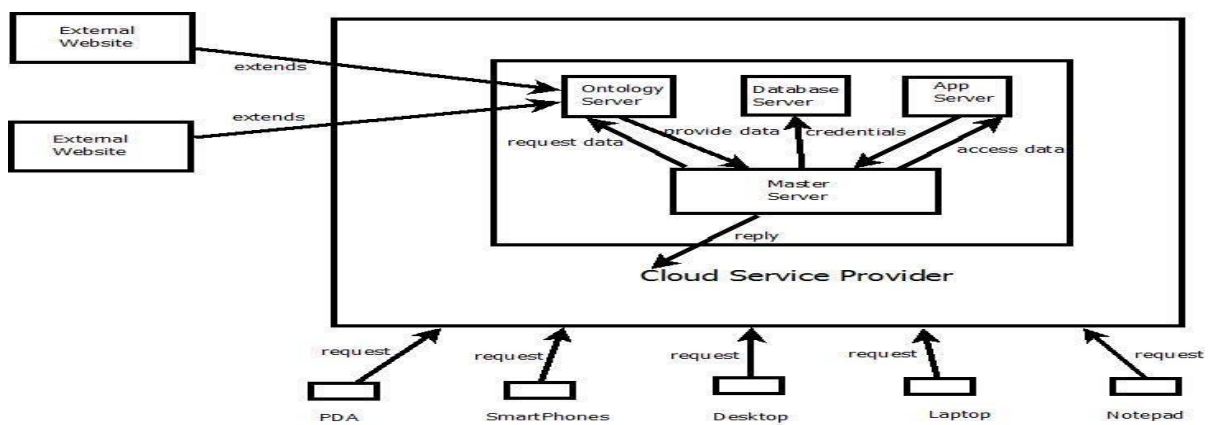

Figure 10 Proposed Architecture for Ontology-based Learning Management System (LMS)

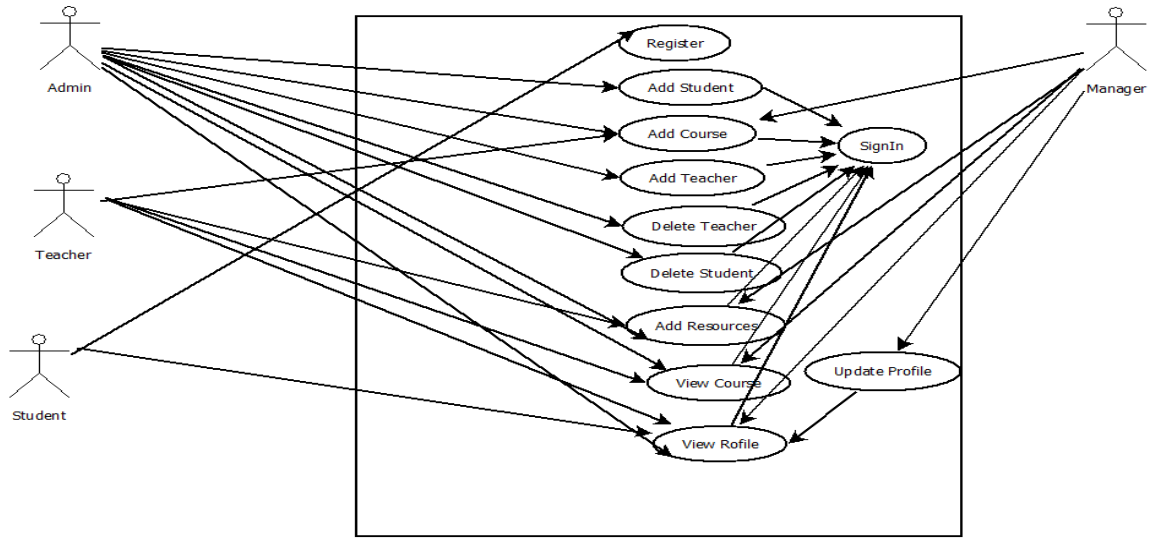

Figure 11 USE CASE

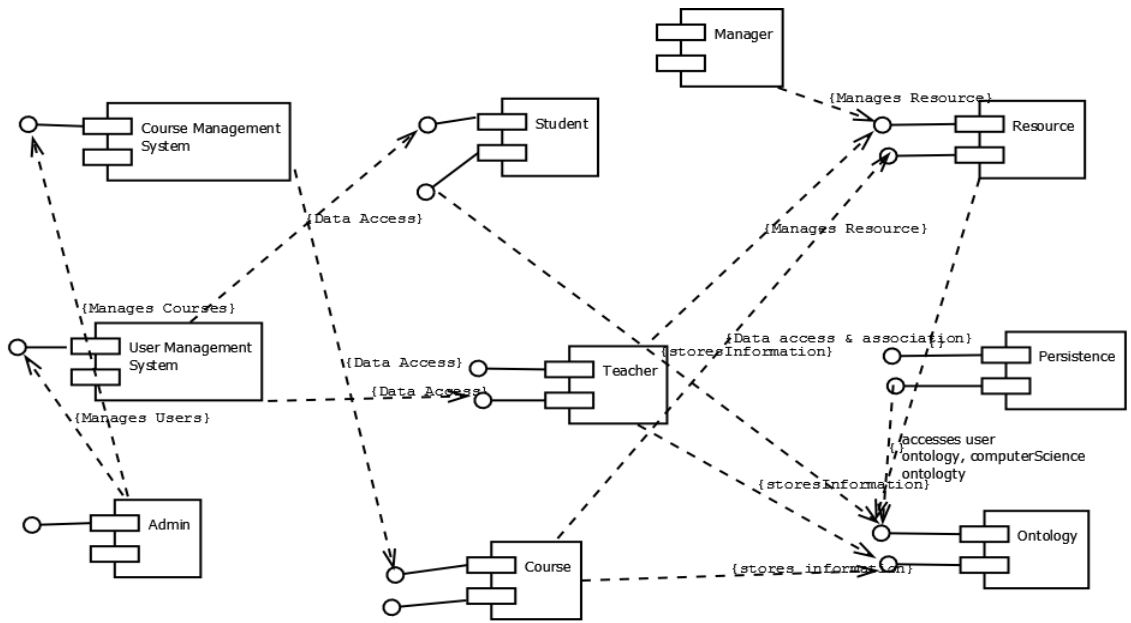

Figure 12 COMPONENT DIAGRAM

```
OWLOntologyManager manager = OWLManager.createOWLOntologyManager();
File file=new File("/mLearning/computerScience.owl");
OWLOntology computerScence=manager.loadOntologyFromOntologyDocument(file);

IRI ontologyIRI = IRI.create("http://www.semanticweb.org/mLearning/ontologies/computerScience");
IRI documentIRI = IRI.create("/mLearning/computerScience.owl");
SimpleIRIMapper mapper = new SimpleIRIMapper(ontologyIRI, documentIRI);
manager.addIRIMapper(mapper);
OWLOntology ontology = manager.createOntology(ontologyIRI);
OWLDataFactory factory = manager.getOWLDataFactory();
OWLClass computerScience = factory.getOWLClass(IRI.create(ontologyIRI + "#ComputerScience"));
```

Figure 13 ComputerScience.owl

```
OWLClass software = factory.getOWLClass(IRI.create(ontologyIRI + "#Software"));
OWLAxiom axiom = factory.getOWLSubClassOfAxiom(computerScience, software);
AddAxiom addAxiom = new AddAxiom(ontology, axiom);
manager.applyChange(addAxiom);
```

Figure 14 Adding super classes and sub classes for Computer Science

```
String base = "http://www.semanticweb.org/mLearning/ontologies/computerScience";
PrefixManager pm = new DefaultPrefixManager(base);
OWLClass communicationManagement=dataFactory.getOWLClass(":Person", pm);
OWLNamedIndividual buffering = dataFactory.getOWLNamedIndividual(":Buffering", pm);
OWLClassAssertionAxiom classAssertion =dataFactory.getOWLClassAssertionAxiom(communicationManagement, buffering);
OWLOntology ontology = manager.createOntology(IRI.create(base));
manager.addAxiom(ontology, classAssertion);
```

Figure 15 Snippet for adding Individuals

```
OWLIndividual x = dataFactory.getOWLNamedIndividual(IRI.create(base + "#xyzTeacher"));
OWLIndividual y = dataFactory.getOWLNamedIndividual(IRI.create(base + "#abcStudent"));
OWLObjectProperty isStudentOf = dataFactory.getOWLObjectProperty(IRI.create(base + "#isStudentOf"));
OWLObjectPropertyAssertionAxiom assertion =dataFactory.getOWLObjectPropertyAssertionAxiom(isStudentOf, abcStudent,xyzTeacher);
AddAxiom addAxiomChange = new AddAxiom(ontology, assertion);
```

Figure 16Snippet for adding objectproperty

```
private static OWLObjectRenderer renderer = new DLSyntaxObjectRenderer();
OWLIndividual y = dataFactory.getOWLNamedIndividual(IRI.create(base + "#abcStudent"));
OWLObjectProperty isPursuing = factory.getOWLObjectProperty(":isPusuing", pm);
LinkedList <OWLObjectProperty>ob=new LinkedList<OWLObjectProperty>();
for (OWLNamedIndividual ind : reasoner.getObjectPropertyValues(y, isPursuing).getFlattened()) {
        ob.add(renderer.render(ind)));
    }
```

Figure 17  DL Query Syntax

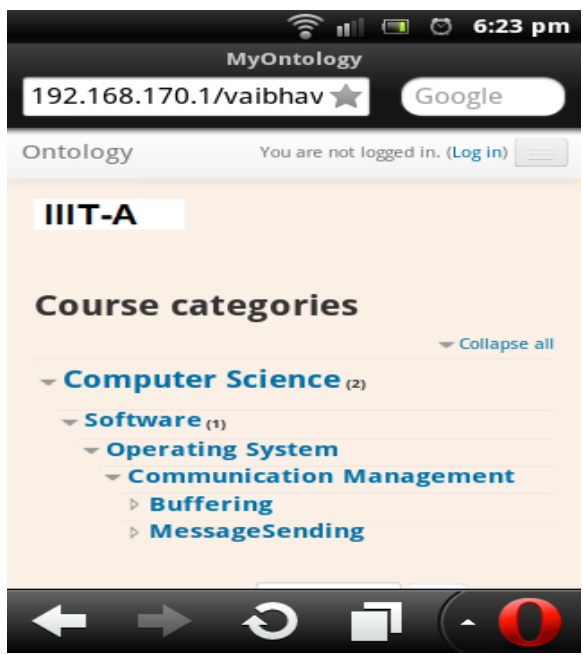

Figure 18 Individuals of Communication Management (App)

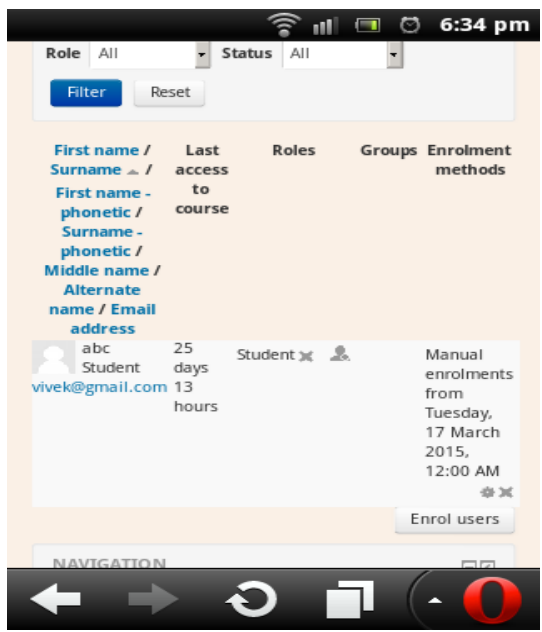

Figure 19  DL Query to extract details of enrolled learner for a subject from Ontology

## EXPERIMENTALRESULT

A website and a mobile app (as shown in Figure18 ) have been developed to implement the proposed an ontological learning management system. The system has made the full use of responsive web design and the multiple learning object repositories can be used to serve multiple formats for many systems and devices. The website has been tested across various browsers and the app has been tested across various devices. The snapshots have been taken on Google Chrome browser and Sony Experia.

The OWL API is used to store the relevant information or data in the ontology. The ontology loading is done at the very first deployment of the application on the server.  The following code snippet as shown in Figure13 is used to load the computerScience.owl ontology.Once the ontology is loaded on the server, we can perform various operations on it. Any course can be added by the admin in the system dynamically with the help of OWL API. Any additions to the system will be reflected in the ontology.

The courses that are under the category of Computer Science such as Software can be  added as subclasses using the snippet as shown in Figure 14. Similarly, the sub-courses such as Programming Language, Operating System, and Software Engineering are added as the sub-classes of Software. The various categories under operating system or theother sub-courses are sub-classes under their respective super-classes. These categories are finally classified as individuals i.e. Communication Management, the sub class of Operating System has individuals such as Buffering, Message Sending, etc using the snippet as shown in Figure15.Similarly, individuals of teacher, admin, student and manager can be added to the user Ontology using Snippet as shown in Figure 16. The snippet represents the associates teacher  i.e. xyzTeacher with a student abcStudent by using the object property "isStudentOf".

The details related to any individuals and classes are extracted from the ontology using DLQuery [54] syntax is shown in Figure17. DLQuery extracts the details from the ontology using 'Manchester Encoding' format which is easy to understand and use. Any student is associated with any course using the object property "enrolledAt" and therefore, this property is used to enlist all the students associated with this course. The result of DLQuery(Figure 17) is shown as a snapshot in Figure19 depicts the same by showing the details of learners/students enrolled for a subject.

The learner goes through a survey where the answers weregiven will decide the learning style. The learning style would decide the delivery mechanism through which knowledge should be provided to the learner.The question consists of four options and all these options correspond to the VARK learning style as shown in Figure 20. The first answer corresponds to kinaesthetic learners whereas the second, third and fourth answer corresponds to visual, read/ write and aural learners respectively. The answers given by learners are used to calculate the score of each learning style and the learning style with the maximum score is taken into consideration [55].

An adaptive quiz has been implemented for self-evaluation of learners. The learners are given two chances to answer any question. They are recommended with proper resources if they fail to answer in both the attempts.The answer is given in thefirst attempt by the learner if found wrong, then hint of an answer to the question is mention to learner shown as in Figure 21 but if the learner fails in giving the correct answer in the second attempt then learner is provided with the proper recommendation to go through the required topic as shown in Figure 22. This quiz system helps the learner in self-assessment and thereby improving the efficiency of Learning Management System (LMS).

The implemented system has granular learning content that is separated from the presentation. The domain ontology, implemented at the elementary level helps to achieve the same. In addition to the basic functionality, of course enrollment, completion rules, competition management, etc., the system also create learning paths for learners and thus provides them with appropriate resources to go through.

CONCLUSION & FUTURE WORK

Learning systems are facing rapid changes with the advent of semantic web technologies, and intelligent learning applications are becoming possible with the development of ontologies. In this paper, Learning Management System (LMS) is implemented using ontologies. The domain ontology facilitates the learner with granular learning content, whereas the task ontology facilitates the system to perform various functionalities. The personalization of data is done by making use of the VARK Learning model along with the domain and task ontology and, therefore, the system is capable of recommending appropriate resources to the learner. Moreover, the adaptive quiz system will help the learners to self-evaluate themselves and recommend right topics to go through. The system is capable of handling the courses of other domains simultaneously as the system is highly scalable, provided that the course added must be mentioned in the ACM Computing Classification System (CSS), for the unification and standardization of ontology.

A feedback system can also be implemented for each and every resource provided by the system. Such system would help to rank the resources. The feedback system can rank the system on the basis of many dimensions. One dimension could be the level of resource whereas the other dimension could be the appropriateness of resources. This would help the system to provide the most appropriate resource for the learner. The Internet of Things (IoT) and their applications in m-learning is just a beginning so that in the future Learning objects can be recommended for learners.

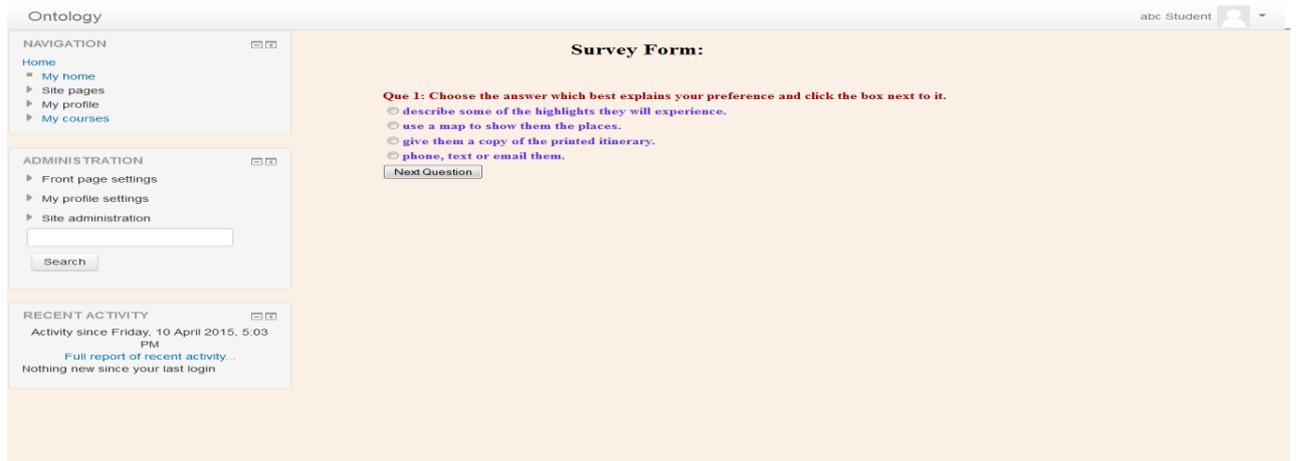

Figure 20 Survey to find Learning Style

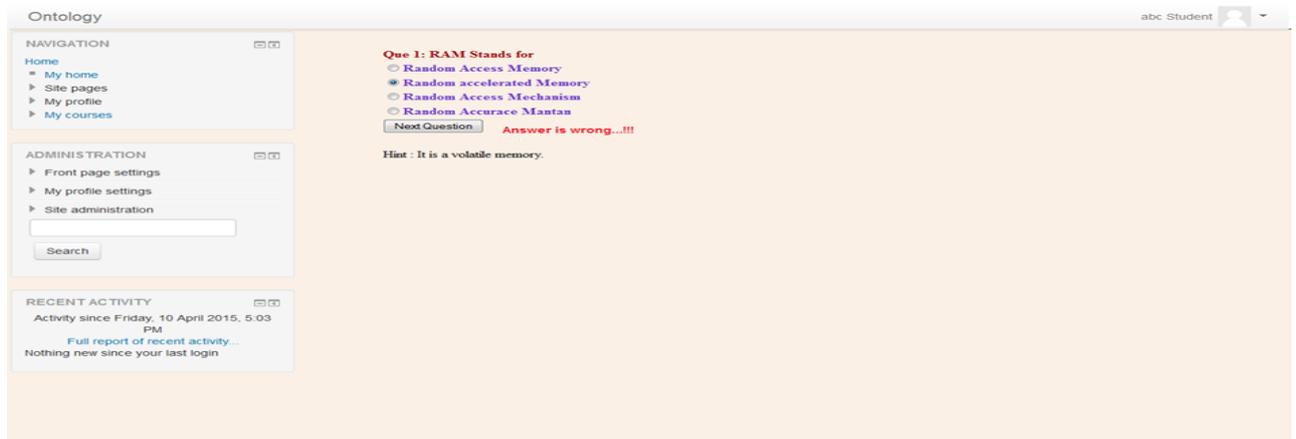

Figure 21 Adaptive quiz system with a hint in the case of first wrong attempt

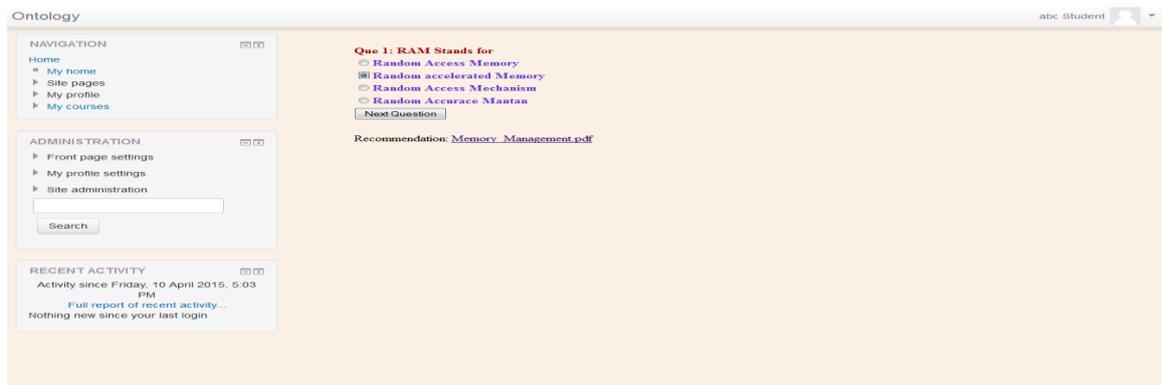

Figure 22 Adaptive quiz system with a recommended resource in the case of second wrong attempt